\documentclass[pra,aps,floatfix,showpacs,showkeys]{revtex4}
\usepackage{graphicx} % Include figure files
\usepackage{epstopdf}
\usepackage{dcolumn}% Align table columns on decimal point
\usepackage{bm}% bold math
\usepackage{amsfonts,amsmath,amssymb%,float
}

\usepackage{bm,dsfont}

\usepackage{bbm}
\usepackage{longtable}
\usepackage[dvips]{epsfig}
\usepackage{hyperref}
\usepackage{listings}
\usepackage{color}
\DeclareMathOperator{\Tr}{Tr}

\newcommand{\unit}{1\!\!1}

\let\phi\varphi

\let\theta\vartheta
\renewcommand{\rho}{\varrho}

\newcommand{\tr}[1]{\operatorname{Tr} #1}

\begin{document}
%\title{Detection of typical bipartite entanglement by generalized local measurements}
\title{Bipartite entanglement detection by local generalized measurements}
	
\author{Maximilian Schumacher, Gernot Alber}
\affiliation{Institut f\"{u}r Angewandte Physik, Technical
University of Darmstadt, D-64289 Darmstadt, Germany}
\date{\today}
\begin{abstract}
Entanglement detection by local measurements, which can possibly be performed by far distant observers, are of particular interest for applications in quantum key distribution and quantum communication. In this paper
sufficient conditions for arbitrary dimensional bipartite entanglement detection based on correlation matrices and joint probability distributions of such local measurements are investigated.  In particular,
their dependence on the nature of the local measurements is explored for typical bipartite quantum states and for measurements involving local orthonormal hermitian operators  bases (LOOs) or generalized measurements based on informationally complete positive operator valued measures of the recently introduced $(N,M)$-type ($(N,M)$-POVMs) \cite{NMPOVM}.
It is shown that symmetry properties of $(N,M)$-POVMs imply that sufficient conditions for bipartite entanglement detection exhibit peculiar scaling properties relating different equally efficient local entanglement detection scenarios. For
correlation-matrix based bipartite local entanglement detection, for example, this has the consequence that LOOs and all informationally complete $(N,M)$-POVMs are equally powerful. With the help of a hit-and-run Monte-Carlo algorithm the effectiveness of  local entanglement detection of typical bipartite quantum states  is explored  numerically. For this purpose Euclidean volume ratios between locally detectable entangled states and all bipartite quantum states are determined.
\end{abstract}
\maketitle
\section{Introduction}
Entanglement \cite{entanglement-general} is not only a characteristic quantum phenomenon of composite quantum systems distinguishing them significantly from classical physical systems but it is also a valuable resource for quantum information science with numerous applications, particularly in quantum key distribution and more generally in quantum communication \cite{entanglement-communication}. 
For two-qubit and qubit-qutrit systems the Peres-Horodecki criterion \cite{Peres,Horodecki} yields a simple necessary and sufficient condition for identifying bipartite entanglement based on the negative partial transpose (NPT) of a quantum state. Although in higher dimensional cases the NPT property is only sufficient and no longer necessary for bipartite entanglement due to the intricate features of bound entanglement \cite{entanglement-general}, establishing the NPT property of a bipartite quantum state is still an important (sufficient) means for proving its entanglement.
However, as it is not known how to implement the operation of partial transposition of an unknown bipartite quantum state  it is of interest to develop quantum measurements which yield sufficient conditions for detecting and thus proving entanglement. Thereby, quantum measurements which can be performed locally, possibly even by far distant observers, are of particular interest for applications in quantum key distribution and quantum communication. 
In this context 
 the natural question arises how does the efficiency of entanglement detection by local measurements depend on the chosen quantum measurements and on the dimensionality of the systems involved for typical quantum states. 
 
In this paper we address this question
 by exploring basic properties of a class of  recently discussed sufficient conditions for arbitrary dimensional bipartite entanglement which are based on correlation matrices and joint probability distributions of local measurements \cite{correlation1,correlation2,correlation3,correlation4}.  In order to explore the dependence of these sufficient conditions on the nature of the local measurements we investigate not only measurements involving local orthonormal hermitian operator bases (LOOs) but also generalized measurements based on informationally complete positive operator valued measures (POVMs) \cite{POVM1,POVM2} of the so called $(N,M)$ type ($(N,M)$-POVMs) \cite{NMPOVM}. These parameterized families of generalized measurements have been introduced recently in order to unify the theoretical description of numerous important classes of quantum measurements, including POVMs, projective measurements involving mutually unbiased bases (MUBs) \cite{MUB}, mutually unbiased measurements (MUMs) \cite{MUM}, symmetric informationally complete measurements (SIC-POVMs) \cite{SIC-POVM1, SIC-POVM2} and their generalizations, so called GSIC-POVMs \cite{GSIC-POVM}. 
In principle, measurements based on informationally complete $(N,M)$-POVMs enable a complete reconstruction of arbitrary quantum states. Therefore, it is expected that these generalized measurements are particularly well suited for detecting bipartite entanglement by local measurements.   
It will be shown that inherent symmetry properties of $(N,M)$-POVMs imply that recently discussed sufficient conditions for bipartite entanglement \cite{correlation1,correlation2,correlation3,correlation4} exhibit peculiar scaling properties which relate different local entanglement detection scenarios. This has the practical consequence that the efficiency of detecting bipartite entanglement by different local measurements, which are related by this scaling property, are identical. In order to explore the efficiency of entanglement detection of typical bipartite states
by local measurements involving these sufficient conditions we determine numerically Euclidean volume ratios between detectable bipartite entangled states and all bipartite quantum states
 %and their dependence on the parameters characterizing
of various dimensions and compare them with the corresponding volume ratios of NPT states. For this purpose we use a hit-and-run Monte Carlo method which has successfully been developed previously for the determination of Euclidean volume ratios of bipartite entangled quantum states \cite{Sauer1,Sauer2}. 
 
 This paper is organized as follows. In Sec. \ref{genmeas} basic properties of $(N,M)$-POVMs,  as introduced originally by Siudzinska \cite{NMPOVM}, are summarized. Furthermore,  the relation is explored between orthonormal hermitian operator bases and informationally complete $(N,M)$-POVMs. These results constitute the basis for our subsequent investigations on the scaling relations between different local entanglement detection scenarios.
In Sec.\ref{correlations} correlation-matrix-based and joint-probability-based sufficient conditions for bipartite entanglement detection by local quantum measurements are discussed within a general framework. Subsequently these results are applied to measurements based on LOOs and on local $(N,M)$-POVMs. As a main result, based on Sec. \ref{genmeas} the resulting scaling relations are derived for LOOs and $(N,M)$-POVMs.
As $(N,M)$-POVMs constitute a unifying framework for the description of many special quantum measurement procedures, including MUMs, SIC-POVMs and GSIC-POVMs, the general results of this section reproduce also previously known results which have been derived for some of these special cases separately. In Sec. \ref{Numerical} numerical results on lower bounds of Euclidean volume ratios of bipartite entangled states are presented for local measurements involving LOOs and $(N,M)$-POVMs and are compared with the corresponding ratios of NPT bipartite quantum states. On the basis of these numerical results on locally detectable entanglement of typical bipartite quantum states the efficiencies of different local entanglement detection procedures can be compared quantitatively.

\section{Informationally complete generalized measurements\label{genmeas}}
In this section basic features of $(N,M)$-POVMs are summarized. This generalization of positive operator valued measures (POVMs) has been introduced recently \cite{NMPOVM}. It generalizes POVMs \cite{POVM1,POVM2} and  describes numerous important quantum measurements procedures in a unified way. In particular, these measurements include
projective measurements with mutually unbiased bases (MUBs) \cite{MUB}, mutually unbiased measurements (MUMs) \cite{MUM}, measurements by symmetric, informationally complete POVMs \cite{SIC-POVM1,SIC-POVM2} and their generalizations, so called GSIC-POVMs \cite{GSIC-POVM}. In this subsection we are particularly interested in informationally complete $(N,M)$-POVMs whose positive (semidefinite) elements span the Hilbert space of all hermitian operators of a quantum system. Starting from the general properties of $(N,M)$-POVMs, summarized in Sec. \ref{generalproperties}, in Sec.\ref{basisPOVM} basic properties of the relation between orthonormal hermitian operator bases and informationally complete $(N,M)$-POVMs are explored. As a main new result of this section it is demonstrated that these relations are characterized by orthogonality preserving linear maps between these operator spaces. In the subspace of hermitian operators, which are orthogonal to the unit operator of the quantum system's Hilbert space, these linear maps act conformally thus scaling all elements of this subspace by the same positive amount.  It is this peculiar scaling property which gives rise to interesting scaling relations for the sufficient bipartite entanglement conditions discussed in Sec. \ref{correlations}.
\subsection{General properties\label{generalproperties}}
Let us consider the $d$-dimensional Hilbert space ${\cal H}_d$ of a quantum system. A $(N,M)$-POVM, described by the $NM$-duple $\Pi=(\Pi_1,\cdots,\Pi_{NM})$, is a set of $(NM)$ positive (semidefinite) operators $\Pi_{i}\geq 0$ ($  i=1,\cdots,NM$) which constitute $N$ different POVMs  formed by the $N$ possible disjoint subsets of cardinality $M$. Therefore, the members $\Pi_i\geq 0$ of such a $(N,M)$-POVM can be identified uniquely by ordered pairs $(\alpha,a)$ with $\alpha$ identifying the particular POVM and $a$ identifying the possible classical measurement result of this particular POVM $\alpha$. Such an identification can be obtained by the bijective map $i: \{(\alpha,a)|\alpha \in \{1,\cdots,N\},~a\in \{1,\cdots,m\}\}\to \{1,\cdots,NM\} $ with
$i(\alpha,a) = (\alpha -1)M + a$, for example. Therefore, each of the $N$ POVMs fulfills the characteristic completeness relations for the possible distinguishable classical measurement results $a \in \{1,\cdots,M\}$, i.e.
\begin{eqnarray}
\sum_{a=1}^M ~\Pi_{i(\alpha,a)} &=&\unit_d
\label{complete}
\end{eqnarray}
for each $\alpha \in \{1,\cdots,N\}$ with
$\unit_{d}$ denoting the unit operator in the Hilbert space ${\cal H}_d$. In addition, 
this set of positive (semidefinite) operators of a $(N,M)$-POVM fulfills the characteristic additional relations \cite{NMPOVM}
\begin{eqnarray}
\Tr\{\Pi_{i(\alpha,a)} \} &=&\frac{d}{M},\label{additional1}\\
\Tr\{\Pi_{i(\alpha,a)} ~\Pi_{i(\alpha,a')} \} &=&x ~\delta_{a,a'} + \nonumber\\
&&(1-\delta_{a,a'})\frac{d-Mx}{M(M-1)},\label{additional2}\\
\Tr\{\Pi_{\alpha,a} ~\Pi_{\beta,b} \} &=& \frac{d}{M^2}
\label{additional3}
\end{eqnarray}
for all $\beta\neq \alpha \in \{1,\cdots,N\}$ and $a,b \in \{1,\cdots,M\}$ if $N\geq 2$. In the degenerate case of a POVM with $N=1$ constraint (\ref{additional3}) is not imposed.
For given values of $(d,N,M)$ each possible value of $x>0$ yields a different $(N,M)$-POVM
and its possible values are constrained by the relation
$d/M^2< x \leq {\rm min}(d^2/M^2,d/M)$. 

A $(N,M)$-POVM $\Pi$ is informationally complete if and only if it contains $d^2$ linearly independent positive operators. As each of the $N$ POVMs involved fulfills the completeness relation (\ref{complete})  this is equivalent to the requirement 
\begin{eqnarray}
(M-1)N + 1 &=& d^2.
\label{dimension}
\end{eqnarray}
There are at
 least four classes of informationally complete $(N,M)$-POVMs corresponding to 
 the possible solutions of (\ref{dimension})\cite{NMPOVM}, namely
 $(N,M) \in \{(1,d^2),~(d+1,d),~(d^2-1,2),~(d-1,d+2)\}$.
The solution $(N,M) = (1,d^2)$ characterizes the special case of a one-parameter family of
generalized symmetric informationally complete positive operator valued measurements (GSIC-POVMs) \cite{GSIC-POVM} parameterized by the parameter $x$. SIC-POVMs correspond to the special case of GSIC-POVMs with $x=1/d^2$.
The solution $(N,M) = (d+1,d)$ describes mutually unbiased
measurements (MUMs) \cite{MUM}, which in the special case of $x=d^2/M^2=d/M=1$ further
 reduce to projective measurements of unit rank with maximal sets of $d+1$ mutually unbiased bases.
In the special case of
a qubit, i.e. $d=2$,
these four possible solutions of (\ref{dimension}) reduce to two cases, namely GSIC-POVMs for $(N,M) = (1,4)$
and MUMs for $(N,M) = (3,2)$.
 
 \subsection{Informationally complete $(N,M)$-POVMs and orthonormal hermitian operator bases\label{basisPOVM}}
 In a Hilbert space ${\cal H}_d$ 
an informationally complete
$(N,M)$-POVM can be expanded in a basis of $d^2$ linearly independent linear operators, say $G=(G_1,\cdots,G_{d^2})^T$, acting on ${\cal H}_d$. These operators can always be chosen as orthonormal hermitian operators with respect to the Hilbert-Schmidt (HS) scalar product
$\langle G_{\mu} | G_{\nu}\rangle_{HS} := \Tr\{G_{\mu}^{\dagger} G_{\nu}\}$ with $G_{\mu}^{\dagger} = G_{\mu}$. They form a basis of the HS-Hilbert space ${\cal H}_{d^2} = ({\rm Span}(G),\langle \cdot | \cdot \rangle_{HS})$ over the field of real numbers which is therefore a Euclidean vector space.
The resulting basis expansion of an arbitrary $(N,M)$-POVM in such an arbitrary orthonormal hermitian basis of linear operators has the general form 
\begin{eqnarray}
\Pi = G^T S
\label{basisexpansion}
\end{eqnarray}
 with $S$ denoting the linear operator which maps ${\cal H}_{d^2}$ into
the possibly higher dimensional HS-Hilbert space ${\cal H}_{NM}$ of hermitian operators. This is a consequence of the dimensional constraint
$NM = d^2 + N-1 \geq d^2$ valid for  informationally complete $(N,M)$-POVMs. The structure of this linear map $S$ and its corresponding
$d^2 \times (NM)$ matrix $S_{\mu,i}$ of real-valued coefficients is significantly constrained by the relations (\ref{additional1}), (\ref{additional2}) and (\ref{additional3}) characterizing $(N,M)$-POVMs.

Let us determine the most general form of the linear operator $S: {\cal H}_{d^2} \to {\cal H}_{NM}$ mapping the HS-Hilbert space ${\cal H}_{d^2}$ into the HS-Hilbert space ${\cal H}_{NM}$ over the field of real numbers containing all possible $(N,M)$-POVMs.
In order to determine this map for an arbitrary $(N,M)$-POVM
we start from (\ref{basisexpansion}) and from the constraints (\ref{additional2}) and (\ref{additional3}). For $N\geq 2$ these constraints can be rewritten in the form
\begin{eqnarray}
&&\left(S^T S\right)_{i(\alpha,a),j(\alpha',a')} =
\Gamma \delta_{i(\alpha,a),j(\alpha',a')} -\label{STS}\\
&& \frac{\Gamma}{M} \left(\bigoplus_{\alpha=1}^N J_{\alpha}\right)_{i(\alpha,a),j(\alpha',a')} +
 \frac{d}{M^2} J_{i(\alpha,a),j(\alpha',a')}\nonumber 
\end{eqnarray}
with 
\begin{eqnarray}
\Gamma &=& \frac{xM^2 -d}{M(M-1)}, 
\label{Gamma1}
\end{eqnarray}
with the $(NM)\times (NM)$ matrix $J$ of all ones, i.e. $J_{i(\alpha,a),j(\alpha',a')}=1$,
and with the $M\times M$ matrices $J_{\alpha}$ of all ones, i.e.
$ \left(J_{\alpha}\right)_{i(\alpha,a),j(\alpha',a')}=\delta_{\alpha  ,\alpha'}$. As outlined in the appendix
the spectrum of the symmetric linear operator $S^TS$ is given by
\begin{eqnarray}
{\rm Sp}(S^TS) &=&\{\Gamma^{(N(M-1))},\frac{dN}{M}^{(1)}, 0^{(N-1)}\}
\label{spectrum}
\end{eqnarray}
with the exponents indicating the multiplicities of the eigenvalues.
In the degenerate case of $N=1$ the zero-eigenvalue disappears.
Thus, according to (\ref{dimension}) for all informationally complete $(N,M)$-POVMs the dimension $D$ of the eigenspace of the non-zero eigenvalues of $S^TS$ is given by
\begin{eqnarray}
D &=& N(M-1) +1 = d^2.
\end{eqnarray}
The spectral representation of this symmetric linear operator is given by
\begin{eqnarray}
\left(S^T S\right)_{i,j} &=& \sum_{\mu=1}^{N(M-1)+1} X_{i,\mu}\Lambda_{\mu} X^T_{\mu, j}
\label{spectral}
\end{eqnarray}
 with 
 \begin{eqnarray}
 \Lambda_1 &=& \frac{dNM}{M^2},~~X_{i,1} = \frac{1}{\sqrt{NM}},\nonumber\\
 \Lambda_{\nu} &=& \Gamma,~~\sum_{a=1}^M X_{i(\alpha,a), \nu} = 0
 \label{eigenvector}
 \end{eqnarray}
for each $\alpha \in \{1,\cdots,N\}$, $i\in \{1,\cdots,NM\}$, $\nu \in \{2,\cdots,N(M-1)+1\}$. 
The $(NM)\times (N(M-1)+1)$ matrix $X_{i,\mu}$ fulfills the orthogonality condition 
\begin{eqnarray}
\sum_{i=1}^{NM}\left(X^T\right)_{\mu, i} X_{i,\nu} &=& \delta_{\mu , \nu}
\label{ortho}
\end{eqnarray} 
for $\mu, \nu \in \{1,\cdots, N(M-1) +1\}$.
Thus, as a consequence of (\ref{dimension}), for an informationally complete $(N,M)$-POVM the most general form of the $d^2\times (NM)$ matrix $S_{\mu,i}$ which is consistent with (\ref{additional2}) and (\ref{additional3}) is given by
\begin{eqnarray}
S_{\mu, i} &=& \sum_{\mu'=1}^{d^2} O^T_{\mu, \mu'} \sqrt{\Lambda_{\mu'}}X^T_{\mu',i}
\label{S1}
\end{eqnarray}
with the arbitrary real-valued orthogonal $d^2\times d^2$ matrix $O$, i.e. $O O^T = O^T O = P_{d^2}$.  Thereby, $P_{d^2}$ denotes the projection operator onto the $(N(M-1)+1)$-dimensional eigenspace of non-zero eigenvalues of the linear operator $S^TS$ acting in the HS-Hilbert space ${\cal H}_{NM}$. Note that in the case of an informationally complete $(N,M)$-POVM this $d^2$-dimensional subspace of ${\cal H}_{NM}$ is isomorphic to the HS-Hilbert space ${\cal H}_{d^2}$.
The additional constraint (\ref{complete}) characterizing any POVM implies the additional relation
\begin{eqnarray}
\unit_{d} &=& \sum_{a=1}^M \Pi_{i(\alpha,a)} = \sqrt{d} \sum_{\mu=1}^{d^2} G_{\mu} O^T_{\mu, 1} 
\label{O1}
\end{eqnarray}
where we have taken into account property (\ref{eigenvector}) of the eigenvectors of $S^TS$. This latter property together with (\ref{O1}) implies that also condition (\ref{additional1}) is fulfilled. All these considerations concerning the spectral representation of $S^TS$ also apply to the degenerate case of $N=1$. However,  in this latter case  all eigenvalues are non-zero.

Therefore, the map $S: {\cal H}_{d^2} \to {\cal H}_{NM}$ as defined by (\ref{S1}) fulfills all requirements defining an informationally complete $(N,M)$-POVM according to the basis expansion (\ref{basisexpansion}).
In a new basis of orthonormal hermitian operators defined by $
\tilde{G}= O G 
$ (cf. (\ref{S1})) the basis expansion (\ref{basisexpansion}) implies
$\Pi = G^T S = \tilde{G}^T \tilde{S}$ with
\begin{eqnarray}
\tilde{S}_{\nu, i} &=& \sqrt{\Lambda_{\nu}}X^T_{\nu,i}
\label{S2}
\end{eqnarray}
and with $\tilde{G}_1 = \unit_d/\sqrt{d}$ and $\Tr\{\tilde{G}_{\nu} \} =0$ for $\nu \in \{2,\cdots,d^2\}$ (cf. (\ref{O1})). In this new basis $\tilde{G}$ it is apparent that according to (\ref{spectrum}) the map $S$ maps ${\cal H}_{d^2}$ injectively onto a $d^2$-dimensional subspace of ${\cal H}_{NM}$ by preserving orthogonality in such a way that all basis operators $\tilde{G}_{\nu}$ with $\nu \in \{2,\cdots,d^2\}$ are mapped conformally onto a $(d^2-1)$-dimensional subspace of ${\cal H}_{NM}$ by stretching the norms of all operators by a factor of $\sqrt{\Gamma}$. As these basis operators are orthogonal to $\tilde{G}_1$ they are characterized by the basis independent property $\Tr\{\tilde{G}_{\nu}\} = 0$ for $\nu \in \{2,\cdots,d^2\}$. It is only the basis operator $\tilde{G}_1$ which is stretched by a different factor, namely $\Lambda_1 = \sqrt{dN/M}$. In the subsequent section it will be demonstrated that this special property relating an arbitrary basis of orthonormal hermitian operators, $G$ or $\tilde{G}$, to an arbitrary informationally complete $(N,M)$-POVM $\Pi$ manifests itself in general scaling relations for sufficient conditions of bipartite entanglement.

 \section{Correlations of separable bipartite quantum states\label{correlations}}
 In this section recently discussed constraints imposed on local correlations  of separable bipartite quantum states are summarized and generalized to $(N,M)$-POVMs. Violating these constraints yields sufficient conditions for bipartite entanglement of arbitrary dimensional quantum systems. In particular, we concentrate on inequalities for the
 $1$-norms of local correlation matrices and of joint local probability distributions. We explore the dependence of these sufficient bipartite entanglement conditions on the types of local measurements performed. As a main result it is demonstrated for correlation-matrix based sufficient conditions that local measurements involving orthonormal hermitian operator (LOO)  bases are as powerful for detecting bipartite entanglement as any locally applied informationally complete $(N,M)$-POVM. This is a consequence of the peculiar scaling property of informationally complete $(N,M)$-POVMs presented in Sec. \ref{basisPOVM}. For
joint bipartite probability distributions of informationally complete local $(N,M)$-POVMs the
scaling property discussed in Sec. \ref{basisPOVM} manifests itself in a more subtle way. It will be shown that in these cases for given values of the dimensions $d_A$ and $d_B$ of Alice's and Bob's quantum systems sufficient conditions for bipartite entanglement of  informationally complete local $(N,M)$-POVMs are still identical provided their properly rescaled $x$-parameters (cf. (\ref{additional2})) agree.

 \subsection{Correlation matrices of general local quantum measurements\label{correlationmatrix}}
  Let us consider arbitrary sets of hermitian operators, say ${\cal A} = \{A_i; i=1,\cdots,N_A\}$ and ${\cal B} = \{B_j; j=1,\cdots,N_B\}$, of two local observers, say  Alice and Bob. These operators are supposed to describe local observables, i.e. hermitian operators, or local POVMs. 
The correlation matrix  of a quantum state $\rho$  associated with these local measurements is defined by
 \begin{eqnarray}
 \left(C({\cal A}, {\cal B}| \rho )\right)_{ij} &=& \Tr\{A_i\otimes B_j\left(\rho - \rho^A\otimes \rho^B\right)\}
 \end{eqnarray}
 with the reduced (local) quantum states of Alice and Bob, $\rho^A = \Tr_B\{\rho\}$ and $\rho^B = \Tr_A\{\rho\}$.
 
 An arbitrary bipartite quantum state $\rho$ is separable if and only if it can be represented in the form of a convex combination of an ensemble of local quantum states  of Alice and Bob, i.e.
 \begin{eqnarray}
 \rho &=&\sum_m\ p_m \rho^A_m \otimes \rho^B_m,
 \end{eqnarray}
 with the probabilities $p_m > 0$ and $\sum_m p_m = 1$. The matrix elements of the correlation matrix of such a general separable quantum state $\rho$ are given by
 \begin{eqnarray}
 \left(C({\cal A}, {\cal B}| \rho )\right)_{ij} &=& \sum_{nm} \left(V_{nm}\right)_{i}
\left(W_{nm}\right)_{j}
 \end{eqnarray}
 with the $N_A$- and $N_B$-dimensional correlation vectors $V_{nm}$ and $W_{nm}$ whose components are given by 
 \begin{eqnarray}
 \left(V_{nm}\right)_i &=&  \sqrt{\frac{p_n p_m}{2}}\Tr\{A_i\rho^A_{n}-A_i\rho^A_{m}\},~(i=1,\cdots,N_A),\nonumber\\
 \left(W_{nm}\right)_j &=&\sqrt{\frac{p_n p_m}{2}} \Tr\{B_j\rho^B_{n}-B_j\rho^B_{m}\},~(j=1,\cdots,N_B).\nonumber\\
 \end{eqnarray}
 With the help of the triangular and the Cauchy-Schwarz inequalities the $1$-norm of this correlation matrix can be upper bounded by the relation \cite{Lai-Luo}
  \begin{eqnarray}
&& ||C({\cal A},{\cal B}|\rho)||_{1} \leq 
 \sum_{nm}  ||V_{nm}||_2 ||W_{nm}||_2 \leq  \label{inequality1}\\
 &&
\sqrt{\sum_{nm} ||V_{nm}||_2^2}\sqrt{\sum_{nm}  ||W_{nm}||_2^2} \leq\sqrt{\Sigma_A \Sigma_B}\nonumber
 \end{eqnarray}
 with
 \begin{eqnarray}
 \sum_{nm}||V_{nm}||_2^2 &=&\sum_{i=1}^{N_A}\sum_{m} p_m \left( \tr\{A_i \rho^A_m \}\right)^2 - \nonumber\\
 &&
 \sum_{i=1}^{N_A}\left( \tr\{A_i \rho^A\}\right)^2,\nonumber\\
  \sum_{nm}||W_{nm}||_2^2 &=&\sum_{j=1}^{N_B}\sum_{m} p_m \left( \tr\{B_j \rho^B_m \}\right)^2 - \nonumber\\
 &&
 \sum_{j=1}^{N_B}\left( \tr\{B_j \rho^B\}\right)^2
 \end{eqnarray}
and with
 \begin{eqnarray}
 \Sigma_A &=& \max_{\sigma^A}\sum_{i=1}^{N_A}\left( \tr\{A_i \sigma^A \}\right)^2-\left( \tr\{A_i \rho^A \}\right)^2,\nonumber\\
 \Sigma_B &=& \max_{\sigma^B}\sum_{j=1}^{N_B}\left( \tr\{B_j \sigma^B \}\right)^2-\left( \tr\{B_j \rho^B\}\right)^2.
 \label{Lambda}
 \end{eqnarray}
%Thereby the relation
% $\sum_{i=1}^{N_A}\sum_{m}p_m \left( \tr\{A_i \rho^A_m\}\right)^2 \leq \max_{\sigma^A}\sum_{i=1}^{N_A}\left( %\tr\{A_i \sigma^A\}\right)^2$ has been as an upper bound.
Accordingly, the upper bounds $\Sigma_A$ and $\Sigma_B$ involve a maximization over all local quantum states of Alice and Bob, $\sigma^A$ and $\sigma^B$.
Inequality (\ref{inequality1}) is a general consequence of bipartite separability of the quantum state $\rho$. It applies to correlation matrices of arbitrary local measurements performed on arbitrary dimensional separable bipartite quantum states. For the special cases of MUMs  and GSIC-POVMs these inequalities have already been derived previously \cite{correlation3}. A violation of inequality (\ref{inequality1}) is a sufficient  condition for bipartite entanglement.

\subsection{Correlation matrices of local orthonormal hermitian operator (LOO) bases\label{correlationmatrixhermitian}}
Let us specialize inequality (\ref{inequality1}) to the correlation matrix of two arbitrary LOO-bases, say $G^A=(G^A_1,\cdots, G^A_{d_A^2})^T$ and $G^B=(G^B_1,\cdots, G^B_{d_B^2})^T$ with   $d_A^2 = N_A$ and $d_B^2 = N_B$ and with $d_A$ and $d_B$ denoting the dimensions of Alice's and Bob's quantum systems.
%the local Hilbert spaces ${\cal H}_A$ and ${\cal H}_B$ of Alice's and Bob's quantum systems. The Hilbert space of the bipartite quantum system is ${\cal H} = {\cal H}_A \otimes {\cal H}_B$. 
For such LOO-bases the right hand sides (\ref{Lambda}) of inequality (\ref{inequality1}) are given by
\begin{eqnarray}
\Sigma_A &=& \max_{\sigma^A} \Tr\{(\sigma^A)^2\} - \Tr\{(\rho^A)^2\} = 1 - \Tr\{(\rho^A)^2\},\nonumber\\
\Sigma_B &=& \max_{\sigma^B} \Tr\{(\sigma^B)^2\} - \Tr\{(\rho^B)^2\} = 1 - \Tr\{(\rho^B)^2\}.\nonumber\\
\end{eqnarray}
Therefore, inequality (\ref{inequality1}) reduces to the form \cite{correlation1,correlation2}
\begin{eqnarray}
||C(G^A,G^B|\rho)||_1^2 \leq (1 - \Tr\{(\rho^A)^2\})(1 - \Tr\{(\rho^B)^2\}).\nonumber
\\
\label{suffcond1}
\end{eqnarray}
A violation of this inequality is a sufficient condition for entanglement of an arbitrary dimensional quantum state $\rho$. This sufficient condition involves the correlation matrix of local measurements and the purities of Alice's and Bob's reduced quantum states.
%It is apparent that due to the presence of the purities of the reduced quantum states of Alice and Bob the sufficient condition of bipartite entanglement resulting from a violation of inequality (\ref{suffcond1}) differs from entanglement-witness-based inequalities which depend on the properties of the bipartite quantum state $\rho$ in a purely linear way.

\subsection{Correlation matrices of local informationally complete $(N,M)$-POVMs\label{corrNMPOVM}}
Let us now specialize the sufficient condition for bipartite entanglement resulting from a violation of inequality (\ref{inequality1}) to two local $(N,M)$-POVMs, say $\Pi^A =(\Pi^A_1,\cdots,\Pi^A_{N_A M_A})$ and $\Pi^B =(\Pi^B_1,\cdots,\Pi^B_{N_B M_B})$, performed by Alice and Bob. For the sake of convenience we introduce the indexing of the POVM elements by the mappings $i(\alpha,a) = (\alpha-1)M_A + a$ and $j(\beta,b) = (\beta -1)M_B +b$.  In the following we concentrate on local $(N,M)$-POVMs which are informationally complete so that relation (\ref{dimension}) applies, i.e. $N_A (M_A-1) +1 = d_A^2$ and $N_B (M_B-1) +1 = d_B^2$, with $d_A$ and $d_B$ denoting the dimensions of the
Alice's and Bob's quantum systems. Each of these local $(N,M)$-POVMs can be expanded in arbitrary LOO-bases, say $G^A$ for Alice and $G^B$ for Bob, i.e. $\Pi^A = (G^A)^T S^A$ and $\Pi^B = (G^B)^T S^B$. Thereby, $S^A$ ($S^B$) denotes the $d_A^2 \times (N_A M_A)$
($d_B^2 \times (N_B M_B)$) matrix of real-valued expansion coefficients for $\Pi^A$ ($\Pi^B$).

With the help of this basis expansion the correlation matrix of local informationally complete $(N,M)$-POVMs can be related to the correlation matrix of these LOO-bases, i.e. 
\begin{eqnarray}
C(\Pi^A,\Pi^B|\rho) &=& (S^A)^T C(G^A,G^B|\rho) S^B.
\label{help21}
\end{eqnarray}
In addition, using the results of Sec.\ref{basisPOVM} (cf. (\ref{ortho}), (\ref{S1}) and (\ref{O1})) it is found that the $1$-norms of both correlation matrices
are related by
\begin{eqnarray}
||C(\Pi^A, \Pi^B|\rho)||_1 &=&||\sqrt{\Lambda^A}C(\tilde{G}^A,\tilde{G}^B|\rho)\sqrt{\Lambda^B}||_1
\end{eqnarray}
with the transformed LOO-bases $\tilde{G}^A = O^A G^A$ and $\tilde{G}^B= O^B G^B$ and with the diagonal matrices
 $\Lambda^A$ and $\Lambda^B$ of the nonzero eigenvalues of $(S^A)^T S^A$ and $(S^B)^T S^B$ (cf. (\ref{spectral})).
In view of (\ref{O1}) for an arbitrary quantum state $\rho$ the correlation matrix fulfills the relations
\begin{eqnarray}
(C(\tilde{G}^A,\tilde{G}^B|\rho))_{1\nu} &=&(C(\tilde{G}^A,\tilde{G}^B|\rho))_{\mu1} =0
\end{eqnarray}
for $\mu \in \{1,\cdots, d_A^2\}$ and $\nu \in \{1,\cdots, d_B^2\}$. As a result the relation between 
the $1$-norms of these correlation matrices obeys the simple scaling relation
\begin{eqnarray}
||C(\Pi^A, \Pi^B|\rho)||_1 = \sqrt{\Gamma_A \Gamma_B}
||C(G^A, G^B|\rho)||_1
\label{scaling10}
\end{eqnarray}
with
\begin{eqnarray}
\Gamma_A &=&\frac{x_A M_A^2 -d_A}{M_A (M_A -1)},~~\Gamma_B = \frac{x_B M_B^2 -d_B}{M_B (M_B -1)}.\label{Gamma}
\end{eqnarray}
With the help of the constraints (\ref{additional1}), (\ref{additional2}) and (\ref{additional3}) characterizing the local $(N,M)$-POVMs also
the relevant upper bounds entering inequality (\ref{inequality1}) can be worked out in a straight forward way yielding the results
\begin{eqnarray}
&&\Sigma_A = \label{upper1}\\
&&\sum_{i=1}^{N_A M_A}\left(\max_{\sigma^A} \left(\Tr\{\Pi^A_{i(\alpha,a)} \sigma^A\}\right)^2 - \left(\Tr\{\Pi^A_{i(\alpha, a)} \rho^A\}\right)^2\right) =\nonumber\\
&&
\Gamma_A
\left(\max_{\sigma^A}\Tr\{(\sigma^A)^2\} - \tr\{(\rho^A)^2\}\right)=
%\nonumber\\&&
\Gamma_A
\left(1 - \tr\{(\rho^A)^2\}\right),\nonumber\\
&&\Sigma_B =\nonumber\\
&&\sum_{j=1}^{N_B M_B} \left(\max_{\sigma^B}\left(\Tr\{\Pi^B_{j(\beta, b)} \sigma^B\}\right)^2 -\left(\Tr\{\Pi^B_{j(\beta, b)} \rho^B\}\right)^2\right) =\nonumber\\
&&\Gamma_B
(1 - \tr\{(\rho^B)^2\}).\nonumber
\label{upper2}
\end{eqnarray}
As a consequence the correlation matrix of informationally complete local $(N,M)$-POVMs obeys the inequality
\begin{eqnarray}
 &&||C(\Pi^A, \Pi^B|\rho)||^2_{1} =
 \Gamma_A \Gamma_B
 ||C(G^A, G^B|\rho)||^2_1 
 \leq\label{inequality2}\\
&&~~~ \Gamma_A \Gamma_B (1 - \tr\{(\rho^A)^2\})(1 - \tr\{(\rho^B)^2\})\nonumber
\end{eqnarray}
for all separable bipartite quantum states. It is apparent that this inequality is identical with inequality (\ref{suffcond1}). Thus, as far as violation of the general inequality (\ref{inequality1}) are concerned, LOO-bases are as powerful in detecting entanglement as local informationally complete $(N,M)$-POVMs. This is a direct consequence of the symmetry of the relations defining $(N,M)$-POVMs and of the resulting scaling relations (\ref{scaling10}) and (\ref{upper1}).
Without addressing the unifying aspects of $(N,M)$-POVMs and their characteristic scaling properties special separate instances of  inequality (\ref{inequality2}), which involve MUMs or GSIC-POVMs as local measurements, have already been  discussed previously  \cite{correlation3}.

\subsection{Joint probability distributions\label{jointprob}}
Based on inequality (\ref{inequality2}) and its violation also sufficient conditions for bipartite entanglement can be derived which involve the joint probability distribution of measurement results of local informationally complete $(N,M)$-POVMs.
%Furthermore, based on this latter inequality also a weaker sufficient condition for bipartite entanglement can be derived. It is independent of the impurities of the reduced density operators of Alice and Bob. Similarly as entanglement witnesses it depends on properties of the bipartite quantum state in a linear way.
For this purpose let us consider
the joint probability distribution 
\begin{eqnarray}
P(\Pi^A, \Pi^B|\rho) &=& \Tr\{(\Pi^A)^T \otimes \Pi^B \rho\}
\end{eqnarray}
resulting from the  local measurements of informationally complete$(N_A,M_A)$- and $(N_B,M_B)$-POVMs, say $\Pi^A$ and $\Pi^B$,
by Alice and by Bob. Applying the triangular inequality to the $1$-norm of this joint probability distribution
%and using (\ref{correlation1}) 
we obtain the inequality
\begin{eqnarray}
||P(\Pi^A, \Pi^B|\rho)||_1 &\leq&
%||\Tr\{E^A \otimes E^B (\rho - \rho^A\otimes \rho^B)\}||_1 + ||\Tr\{E^A \otimes E^B \rho^A\otimes \rho^B\}||_1 = 
||C(\Pi^A, \Pi^B|\rho)||_1+ \\
&&||\Tr\{(\Pi^A)^T \otimes \Pi^B \rho^A\otimes \rho^B\}||_1.\nonumber
\end{eqnarray}
From the defining properties of informationally complete $(N,M)$-POVMs (cf. Sec.\ref{basisPOVM}) we obtain the relation
\begin{eqnarray}
&&||\Tr\{(\Pi^A)^T \otimes \Pi^B \rho^A\otimes \rho^B\}||_1 = \sqrt{U_A U_B},\nonumber\\
U_A&=&\Tr\{(\rho^A)^2\}\Gamma_A + \frac{d_A N_A/M_A - \Gamma_A}{d_A}
,\nonumber\\
U_B&=&
\Tr\{(\rho^B)^2\}\Gamma_B
+ \frac{d_B N_B/M_B - \Gamma_B}{d_B}.\label{Us}
\end{eqnarray}
Thus, 
using the upper bounds on the correlation matrix (\ref{upper1})  we obtain the result
\begin{eqnarray}
&&||P(\Pi^A, \Pi^B|\rho)||_1 \leq \sqrt{\Sigma_A \Sigma_B} + \sqrt{U_A U_B}.
%c&=&
%\sqrt{
%\frac{x_A M_A^2 -d_A}{M_A (M_A -1)}\left(1 - \tr\{(\rho^A)^2\}\right)},\nonumber\\
%d&=&\sqrt{
%\frac{x_B M_B^2 -d_B}{M_B (M_B -1)}(1 - \tr\{(\rho^B)^2\})}
\label{inequality4}
\end{eqnarray}
For separable quantum states $\rho$ this inequality constrains the joint probabilities of the local measurement results of Alice and Bob and relates them to the purities of the reduced quantum states of Alice and Bob, i.e. $\Tr\{(\rho^A)^2\}$ and  $\Tr\{(\rho^B)^2\}$. 
A violation of this inequality by a quantum state $\rho$ yields a sufficient condition for its bipartite entanglement. 
With the help of the general inequality 
\begin{eqnarray}
\sqrt{\Sigma_A \Sigma_B}+\sqrt{U_A U_B} &\leq&\sqrt{\Sigma_A + U_A} \sqrt{\Sigma_B + U_B}
\label{help}
\end{eqnarray}
a new upper bound can be derived on the right hand side of (\ref{inequality4}) yielding the constraint \cite{NMPOVM}
\begin{eqnarray}
||P(\Pi^A, \Pi^B|\rho)||_1 &\leq& 
\sqrt{\Sigma_A + U_A} \sqrt{\Sigma_B + U_B}
\label{inequality5}
\end{eqnarray}
with
\begin{eqnarray}
\Sigma_A + U_A &=&\Gamma_A \left(1 - \frac{1}{d_A}\right) + \frac{N_A}{M_A},
\nonumber\\
\Sigma_B + U_B &=&\Gamma_B \left(1 - \frac{1}{d_B}\right) + \frac{N_B}{M_B}
\label{inequality6}
\end{eqnarray}
where $N_A$ and $N_B$ are related to the corresponding dimensions $d_A$ and $d_B$ by the general condition (\ref{dimension}) of informationally complete POVMs. A violation of this inequality by a quantum state $\rho$ again yields a sufficient condition for its entanglement. However, in view of (\ref{help}) this sufficient condition for bipartite entanglement is generally weaker than the sufficient condition based on a violation of inequality (\ref{inequality4}). It is apparent that inequality (\ref{inequality5}) is independent of the purities of the reduced density operators of Alice and Bob. 
%Thus, the resulting sufficient condition of bipartite entanglement based on a violation of this inequality is
%in the spirit of similar entanglement-witness-based sufficient conditions for entanglement which depend on quantum states only in a linear way. 
For special cases, namely for SIC-POVMs \cite{correlation4} and for GSIC-POVMs \cite{Lai-Li}, inequality (\ref{inequality5}) has already been derived recently.

The sufficient conditions for entanglement resulting from violations of (\ref{inequality4}) or (\ref{inequality5}) for the
joint probability distributions of local $(N,M)$-POVMs also exhibit scaling properties. However, they differ from the scaling properties of the correlation matrix as discussed in Sec. \ref{correlationmatrix}. 
In order to derive these scaling relations let us first of all consider the common left hand side of inequalities (\ref{inequality4}) and (\ref{inequality5}). Expanding the local $(N,M)$-POVMs in arbitrary LOO-bases $G^A$ and $G^B$ according to (\ref{basisexpansion}), i.e. $\Pi^A = (G^A)^T S^A$ and $\Pi^B= (G^B)^T S^B$, we find with the help of the results of Sec.
\ref{basisPOVM} the relation
\begin{eqnarray}
||P(\Pi^A,\Pi^B|\rho)||_1 &=&||\sqrt{\Lambda^A} P(\tilde{G}^A,\tilde{G}^B|\rho) \sqrt{\Lambda^B}||_1
\end{eqnarray}
with the diagonal matrices $\Lambda^A$ and $\Lambda^B$ of the nonzero eigenvalues of the matrices $(S^A)^T S^A$ and $(S^B)^T S^B$ and with the new LOO-bases
$\tilde{G}^A = O^A G^A$ and $\tilde{G}^B = O^B G^B$ (cf. Eqs. (\ref{S1} and (\ref{O1})). According to (\ref{eigenvector}) the common factors 
$\gamma_A= \left(d_A(d_A-1)\right)/\left(M_A(M_A-1)\right)$ and $\gamma_B= \left(d_B(d_B-1)\right)/\left(M_B(M_B-1)\right)$
can be extracted from the eigenvalues entering the diagonal matrices $\Lambda^A$ and $\Lambda^B$, i.e. 
\begin{eqnarray}
(\Lambda^A)_{11} &=&
\gamma_A
%\frac{d_A(d_A+1)}{M_A(M_A-1)}
(d_A+1),\nonumber\\
(\Lambda^A)_{\nu \nu} &=& \Gamma_A = \gamma_A
%\frac{d_A(d_A-1)}{M_A(M_A-1)}
\frac{d_A \tilde{x}_A -1}{d_A-1},\nonumber\\
(\Lambda^B)_{11} &=&\gamma_B
%\frac{d_B(d_B-1)}{M_B(M_B-1)}
(d_B+1),\nonumber\\
(\Lambda^B)_{\mu \mu} &=& \Gamma_B = \gamma_B
%\frac{d_B(d_B-1)}{M_B(M_B-1)}
\frac{d_B \tilde{x}_B -1}{d_B-1}
\end{eqnarray}
for $\nu \in \{2,\cdots,N_A(M_A-1)+1\}$ and 
$\mu \in \{2,\cdots,N_B(M_B-1)+1\}$
with the rescaled parameters 
\begin{eqnarray}
\tilde{x}_A&=& \frac{x_AM_A^2}{d_A^2},~~\tilde{x}_B = \frac{x_BM_B^2}{d_B^2}.
\label{scaledparameter}
\end{eqnarray}
The same factors $\gamma_A$ and $\gamma_B$  can be extracted also from the quantities $U_A$, $U_B$, $\Sigma_A$ and $\Sigma_B$ of (\ref{upper1}) and (\ref{Us}) entering the right hand sides of inequalities (\ref{inequality4}) and (\ref{inequality5}). Consequently these inequalities 
can be rewritten in the rescaled form
\begin{eqnarray}
||\sqrt{\frac{\Lambda^A}{\gamma_A}} P(\tilde{G}^A,\tilde{G}^B|\rho) \sqrt{\frac{\Lambda^B}{\gamma_B}}||_1 &\leq& \sqrt{\frac{U_A}{\gamma_A}}
\sqrt{\frac{U_B}{\gamma_B}},\label{finalinequality}\\
||\sqrt{\frac{\Lambda^A}{\gamma_A}} P(\tilde{G}^A,\tilde{G}^B|\rho) \sqrt{\frac{\Lambda^B}{\gamma_B}}||_1 &\leq& \sqrt{1+\tilde{x}_A}\sqrt{1+\tilde{x}_B}.\nonumber
\end{eqnarray}
It is apparent that for given dimensions $d_A$ and $d_B$ of Alice's and Bob's quantum systems these inequalities depend only on the scaled parameters  $\tilde{x}_A$ and $\tilde{x}_B$ of Eqs.  (\ref{scaledparameter}) characterizing Alice's and Bob's local $(N,M)$-POVMs. This implies that for given dimensions $d_A$ and $d_B$ testing for violations of the inequalities
(\ref{finalinequality}) yields the same results for
informationally complete $(N,M)$-POVMs with parameters $(M_A, x_A, M_B, x_B)$ and with parameters $(M'_A, x'_A=d_A^2 x_A/(M'_A)^2,M'_B, x'_B=d_B^2 x_B/(M'_B)^2)$. Thus, according to (\ref{finalinequality}) the sufficient conditions for bipartite entanglement involving local MUMs with parameters $(M_A=d_A,x_A,M_B=d_B,x_B)$, for example, are identical with the sufficient conditions involving local GSIC-POVMs with $(M'_A=d_A^2, x'_A=x_A/M'_A,M'_B=d_B^2,x'_B=x_B/M'_B)$.

\section{Numerical results\label{Numerical}}
In this section numerical results are presented exploring typical statistical features of bipartite entanglement which can be detected by violations of inequalities (\ref{inequality2}), (\ref{inequality4}) and (\ref{inequality5}) which yield sufficient conditions for bipartite entanglement. Based on these sufficient conditions we determine lower bounds on the Euclidean volume ratios between entangled bipartite states and all quantum states for different dimensions $d_A$ and $d_B$ of Alice's and Bob's quantum systems.

Starting from the $(d_A\times d_B)$-dimensional Hilbert space ${\cal H}_{d_A \times d_B}$
describing Alice's and Bob's joint quantum system one can construct
the $(d_A\times d_B)^2$-dimensional Hilbert space ${\cal H}_{(d_A \times d_B)^2}$ of hermitian linear operators acting on elements of ${\cal H}_{d_A \times d_B}$.
This is a Hilbert space over the field of real numbers equipped with the
Hilbert-Schmidt (HS) scalar product  $\langle A | B\rangle_{HS} := \Tr_{AB}\{A^{\dagger} B\}$ for $A,B\in {\cal H}_{(d_A \times d_B)^2}$. Thus, it is a Euclidean vector space. On this Euclidean vector space volumes of convex sets of linear hermitian operators and thus also of quantum states $\rho \geq 0$ can be defined in a natural way. The Euclidean volumes of quantum states can be estimated numerically with the help of Monte-Carlo methods. We have developed a hit-and-run Monte Carlo algorithm \cite{Sauer1} for estimating the volumes of convex sets of quantum states according to this Euclidean measure. Hit-and-run Monte-Carlo methods have been introduced originally by Smith \cite{Smith}. They take advantage of a random walk inside a convex set in order to generate efficiently a uniform distribution over this convex set by iteration so that this distribution eventually becomes independent of the starting point \cite{Lovasz}. 

We have sampled $N=10^8$ bipartite quantum states randomly for different values of $d_A$ and $d_B$ with the help of this recently developed hit-and-run Monte-Carlo algorithm \cite{Sauer1,Sauer2} in order to determine lower bounds on Euclidean volumes of detected entangled states. In Table \ref{Tab1} the obtained lower bounds of the ratios $R$ between the Euclidean volumes of entangled states and all bipartite quantum states are presented together with their numerical errors. These errors have been estimated with the help of the procedure described in Ref.\cite{Sauer1}. Thereby,  for each pair of dimensions of Alice' and Bob's quantum systems $(d_A,d_B)$ four different lower bounds are presented, namely the volume ratio resulting from bipartite NPT states ($R_{NPT}$) and the volume ratios resulting from bipartite entangled states detectable by violations of inequalities (\ref{inequality5}) ($R_{SIC1}$),  (\ref{inequality4}) ($R_{SIC2}$) and (\ref{suffcond1}) ($R_{LOO}$) which is equivalent with (\ref{inequality2}). Thereby, the ratios $R_{SIC1}$ and $R_{SIC2}$ describe bipartite entanglement detectedable by local informationally complete SIC-POVMs performed by Alice and Bob with the parameters $M_A=d_A^2, x_A = 1/d_A^2, M_B=d_B^2, x_B = 1/d_B^2$ corresponding to the scaled parameters $\tilde{x}_A = \tilde{x}_B =1$. It is apparent from Table \ref{Tab1} that for $d_A=d_B$ the values of $R_{SIC1}$ are consistent with the recently obtained values of Ref. \cite{correlation4}.

The ratios $R_{LOO}$ describe bipartite entanglement detectable by measurements involving LOO-bases. From these results it is apparent that all lower bounds on the ratios $R$ based on local measurements are always smaller than the ratios $R_{NPT}$ of bipartite NPT states. Furthermore, with increasing dimensions of Alice's and Bob's quantum systems the ratios between bipartite entangled states, which are lower bounded by the ratios of NPT states, and all quantum states rapidly approach unity within our  numerical accuracy. However, the lower bounds on these ratios detectable by local measurements do not reflect this tendency and even tend to decrease with increasing dimensions of Alice's and Bob's quantum systems. Nevertheless, as expected on the basis of our discussion in Sec. \ref{correlations} these latter lower bounds always fulfill the relation $R_{SIC1} \leq R_{SIC2}\leq R_{LOO}$ consistent with our results (\ref{inequality2}), (\ref{inequality4}), (\ref{help}) and (\ref{inequality5}).

\begin{table}
  \centering
\begin{tabular}{ |c|c|c|c|c| }
\hline
$(d_A,d_B)$ & $R_{NPT}$ &  $R_{SIC1}$ & $R_{SIC2}$&$R_{LOO}$\\
\hline
$(2,2)$ & $0,75784$ &  $0,67060$&$0,67947 $&$0,68860$\\
&$\pm 1,7(4)$&$\pm 2,2(4)$&$\pm 2,1(4)$&$\pm 2,1(4)$\\
\hline
$(2,3)$&$0,97303$ &  $0,39732$&$0,42998
$&$0,43853$\\
&$\pm 7(5)$ &  $\pm 5,6(4)$&$\pm 5,5(4)
$&$\pm 5,5(4)$\\
\hline
$(2,4)$  &$0,998696$ & $0,02710
$  &$0,04361$& $0,04504$\\
  &$\pm 1,6(5)$ & $
\pm 2,7(4)$ & $\pm 3,4(4)$& $\pm 3,5(4)$\\
\hline
$(3,3)$  &$0,999895
$ & $0,75680$& $0,75754$& $0,76364$\\
  &$\pm 4(6)$ & $\pm 8,2(4)$& $\pm 8,2(4)$& $\pm 8,1(4)$\\
\hline
$(3,4)$  &$1$ & $0,3605$ &$0,3742$ & $0,3795$\\
  &$\pm 0$ & $\pm 1,8(3)$ &$\pm 1,8(3)$ & $\pm 1,8(3)$\\
\hline
$(4,4)$  &$1$ & $0,6378$ & $0,6380
$ & $0,6419$\\
  &$\pm 0$ & $\pm 7,7(3)$ & $
\pm 7,7(3)$ & $\pm 7,7(3)$\\
\hline
\end{tabular}
\caption{Lower bounds on volume ratios between  entangled and all bipartite quantum states for different dimensions $d_A$ and $d_B$ of Alice's and Bob's quantum systems: Bipartite NPT states ($R_{NPT}$), bipartite states detectable by violations of (\ref{inequality5}) ($R_{SIC1}$), of (\ref{inequality4})  ($R_{SIC2}$) of (\ref{suffcond1}) or equivalently of (\ref{inequality2}) ($R_{LOO}$).
The numbers in brackets after the estimated numerical errors \cite{Sauer1} indicate the relevant powers of $10^{-1}$. 
}
\label{Tab1}
\end{table}

\begin{figure}
\includegraphics[width=1.0\linewidth,height=0.4\textheight]{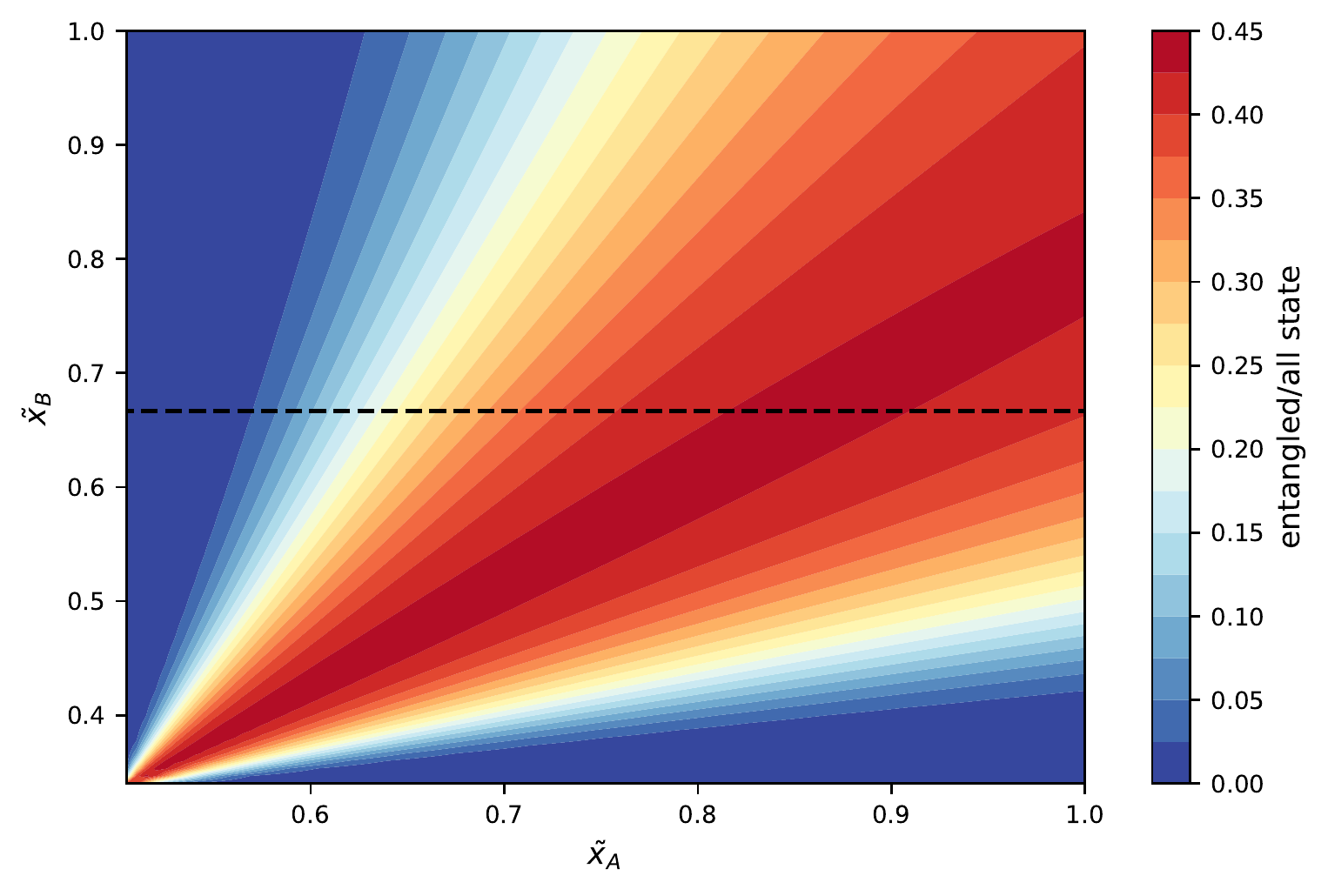}
\caption{Dependence of 
volume ratios between  entangled and all bipartite quantum states on the scaled parameters $\tilde{x}_A $ and $\tilde{x}_B$ of Alice's and Bob's informationally complete local $(N,M)$-POVMs as obtained from a violation of inequality (\ref{inequality5}) for $(d_A,d_B) = (2,3)$: The horizontal dashed line indicates the upper bound for $\tilde{x}_B$ for $(N,M)$-POVMs with $M_B=2$. All $(N,M)$-POVMs with  these values of $d_A$ and $d_B$ yield the same results.
}
\label{Fig1}
\end{figure}

\begin{figure}
\includegraphics[width=1.0\linewidth,height=0.4\textheight]{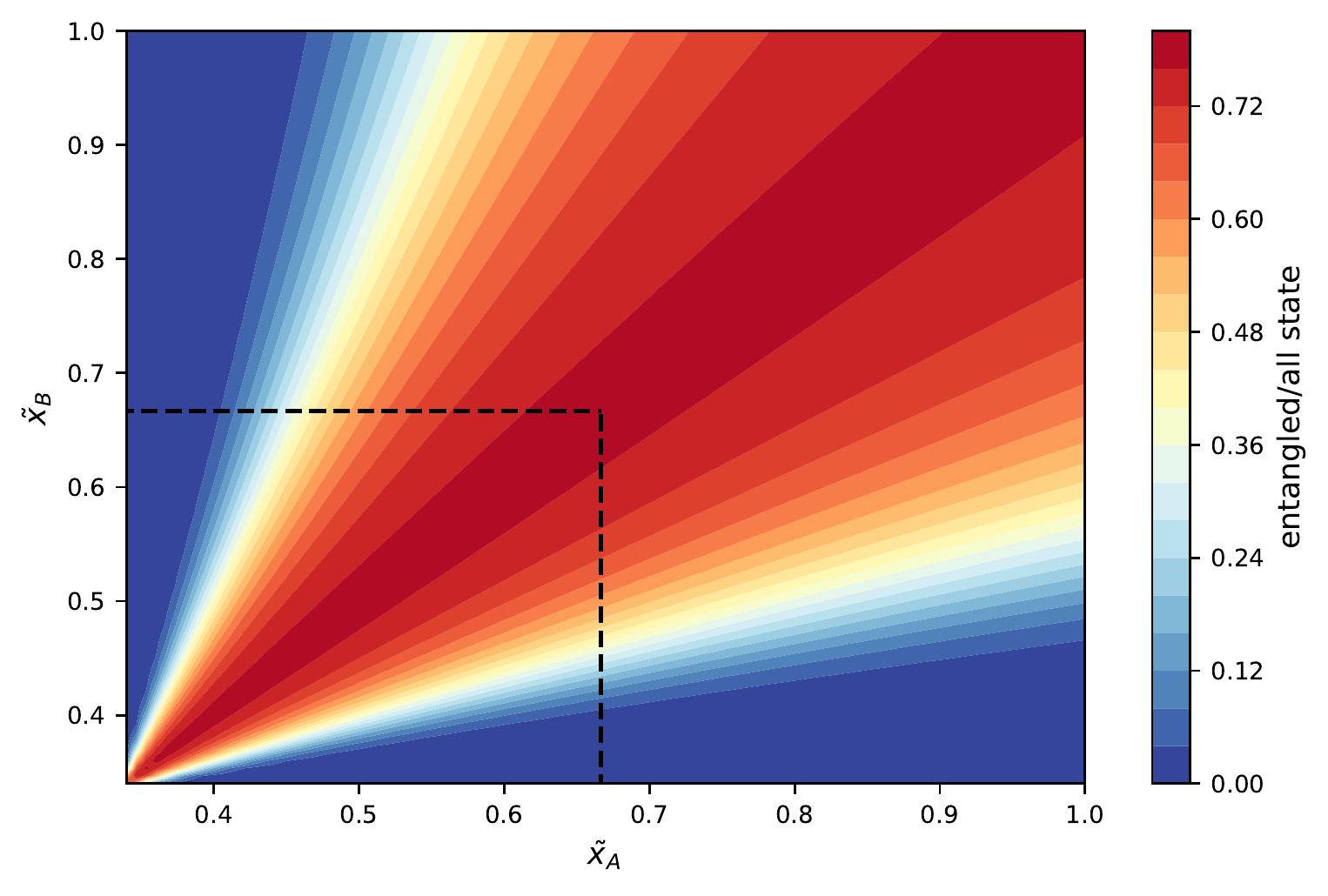}
\caption{Dependence of 
volume ratios between  entangled and all bipartite quantum states on the scaled parameters $\tilde{x}_A $ and $\tilde{x}_B$ of Alice's and Bob's informationally complete local $(N,M)$-POVMs as obtained from a violation of inequality (\ref{inequality5}) for $(d_A,d_B) = (3,3)$: The vertical and horizontal dashed lines indicate the upper bounds for $\tilde{x}_A$ and $\tilde{x}_B$ for $(N,M)$-POVMs with $M_A=2$ and $M_B=2$. All $(N,M)$-POVMs with  these values of $d_A$ and $d_B$ yield the same results.
}
\label{Fig2}
\end{figure}

In Figs. \ref{Fig1} and \ref{Fig2} for a qubit-qutrit and a qutrit-qutrit system the volume ratios of bipartite entangled states detectable by violations of inequality (\ref{inequality5}) and their dependence on  the scaled parameters $\tilde{x}_A $ and $\tilde{x}_B$ of Alice's and Bob's local informationally complete $(N,M)$-POVM measurements (cf. \ref{scaledparameter}) are depicted. These results are based on a statistical ensemble of $10^7$ randomly sampled bipartite quantum states. According to the scaling properties discussed in Sec.\ref{jointprob} (cf. (\ref{finalinequality}) and the discussion of Sec. \ref{genmeas} these results describe the volume ratios for all local $(N,M)$-POVMs of Alice and Bob with scaled parameters in the maximally allowed ranges $1/d_A < \tilde{x}_A \leq {\min}(1,M_A/d_A)$ and $1/d_B \leq \tilde{x}_B \leq {\min}(1,M_B/d_B)$. For $d_A=2$ the possible informationally complete $(N,M)$-POVMs are characterized by the values $(N_A,M_A) \in \{(1,4), (3,2)\}$ so that the corresponding possible range of scaled $x$-parameters is given by $1/2 < \tilde{x_A} \leq 1$ for all these possible $(N,M)$-POVMs. For $d_B=3$ the possible informationally complete $(N,M)$-POVMs are characterized by the values $(N_B,M_B) \in \{(1,9), (2,5), (4,3), (8,2)\}$ so that the corresponding possible range of scaled $x$-parameters is given by $1/3 < \tilde{x}_B \leq 1$ for $M_B\geq 3$ and $1/3< \tilde{x}_B\leq 2/3$ for $M_B=2$. The black dashed straight horizontal line in Fig. \ref{Fig1} marks this upper bound of $2/3$ for $\tilde{x}_B$ for the case $M_B=2$ and  $d_B=3$. In the case depicted in Fig. \ref{Fig2} we have $d_A=d_B=3$, and the corresponding upper bounds for $(N,M)$-POVMS with $M_A=2$ or $M_B=2$ are also indicated by black dashed straight vertical and horizontal lines.

\section{Conclusions}
Basic properties of sufficient conditions for arbitrary dimensional bipartite entanglement detection based on correlation matrices and joint probability distributions of local measurements have been investigated. These sufficient conditions are interrelated by a series of inequalities, and thus differ in their effectiveness of detecting bipartite entanglement by violations of these inequalities.
The dependence of these sufficient conditions on the nature of the local measurements has been explored for measurements involving LOOs and for generalized measurements based on informationally complete $(N,M)$-POVMs.

As a main result it has been shown that inherent symmetry properties of $(N,M)$-POVMs imply that sufficient conditions for bipartite entanglement exhibit peculiar scaling properties which relate different local measurement scenarios. This has the practical consequence that the efficiency of detecting bipartite entanglement by different local measurements related by this scaling are identical. If correlation matrices are used for local bipartite entanglement detection it turns out that in view of these scaling properties all $(N,M)$-POVMs and LOOs are equally powerful. However, if bipartite entanglement detection  is based on local joint probability distributions the scaling relations between different $(N,M)$-POVMs are more subtle. Nevertheless, for given dimensions of Alice's and Bob's quantum systems $d_A$ and $d_B$, for all local $(N,M)$-POVMs the dependence on their scaled parameters $\tilde{x}_A$ and $\tilde{x}_B$ is identical.

In order to address also the question to which extent local measurements are capable of detecting typical bipartite entanglement we have
also determined numerically Euclidean volumes ratios between locally detectable entangled and all bipartite quantum states for different dimensions of Alice's and Bob's quantum systems. For this purpose we have used a previously developed hit-and-run Monte-Carlo algorithm. 
Comparison of these results with the corresponding volumes ratios of NPT bipartite states demonstrates quantitatively to which extent these local entanglement detection procedures underestimate the presence of bipartite entanglement.
In particular, they support the conjecture that with increasing dimensions of Alice's and Bob's quantum systems the efficiency of these local entanglement detection procedures decreases.
 
\begin{acknowledgments}
Useful discussions with A. Sauer and J. Z. Bernad are acknowledged. This research was supported by the Deutsche Forschungsgemeinschaft (DFG) -- SFB 1119 -- 236615297.
\end{acknowledgments}

\appendix
\section{Appendix}
In this appendix the derivation of the spectral representation of the linear operator $S^T S$ of (\ref{STS}) is 
outlined.

We start from the expansion of an informationally complete $(N,M)$-POVM $\Pi=(\Pi_1,\cdots,\Pi_{NM})$ in an orthonormal hermitian operator basis $G=(G_1,\cdots,G_{d^2})^T$, i.e. $\Pi = G^T S$, as discussed in Sec.\ref{basisPOVM}. Defining $i(\alpha,a) := (\alpha-1)M + a \in \{1,\cdots, NM\}$ with $\alpha \in \{1,\cdots, N\}$ and $a \in \{1,\cdots, M\}$ the defining properties of a $(N,M)$-POVM (\ref{additional2}) and (\ref{additional3}) can be rewritten in the form (\ref{STS}) for $N\geq 2$. In the degenerate case of $N=1$ Eq.(\ref{STS}) has to be replaced by
\begin{eqnarray}
(S^T S)_{i(1,a),j(1,a')} &=&
\Gamma \delta_{i(1,a),j(1,a')}+\nonumber\\
&&\frac{d-Mx}{M(M-1)}(J_1)_{i(1,a),j(1,a')}
\label{N=1}
\end{eqnarray}
with the $M\times M$ matrix of all ones, i.e. $(J_1)_{i(1,a),j(1,a')}=1$ for $a,a' \in \{1,\cdots,M\}$.

Let us first consider cases with $N\geq 2$.
It is apparent from (\ref{STS}) that for $N \geq 2$ 
\begin{eqnarray}
S^TS J &=& \frac{dN}{M} J
\end{eqnarray}
so that $X_{i,1} = 1/\sqrt{NM}$ with components $i \in \{1,\cdots, NM\}$ is normalized eigenvector with eigenvalue $\Lambda_1 = dN/M$. As $(S^T S)_{i,j}$ is a symmetric real-valued matrix it can be diagonalized and all its other eigenvectors have to be orthogonal to this particular eigenvector. 
From the form of (\ref{STS}) it is also apparent that all orthonormal vectors
$X_{i,\nu}$ with $\sum_{a=1}^M X_{i(\alpha,a), \nu} = 0$ have the same eigenvalue $ \Gamma$ (cf. (\ref{Gamma1})) because $(J_{\alpha}X)_{i,\nu} = (JX)_{i,\nu} =0$ for $\alpha \in \{1,\cdots,N\}$. For each $\alpha$ it is possible to construct $(M-1)$ such vectors. Therefore, the eigenvalue $\Gamma$ is $N(M-1)$-fold degenerate. There are additional $(N-1)$ linear independent orthonormal vectors $X_{i(\alpha,a),\nu}$  with the property $(\oplus_{\alpha=1}^N J_{\alpha}X)_{i,\nu}=M X_{i,\nu}$ and $(JX)_{i,\nu} =0$. They all have the same eigenvalue $0$. All in all, there are $1+ N(M-1) + (N-1) = NM$ orthogonal eigenvectors and we obtain the spectrum (\ref{spectrum}) of $S^TS$ with its degeneracies and the spectral representation (\ref{spectral}) with the characteristic properties of the eigenvectors (\ref{eigenvector}) and (\ref{ortho}).

Analogous arguments can be used in the degenerate case of $N=1$. In this case the spectrum of $S^TS$ is given by
\begin{eqnarray}
{\rm Sp}(S^T S) &=&\{\Gamma^{(M-1)}, \frac{d}{M}^{(1)}\}
\end{eqnarray}
and the eigenvalue $0$ no longer appears in the spectrum. Relations (\ref{eigenvector}) and (\ref{ortho}) are still valid for the eigenvectors. Furthermore, such a POVM is informationally complete if and only if $M=d^2$.

\end{document}